\documentclass[lettersize,journal]{IEEEtran}
\usepackage{amsmath,amsfonts}
\usepackage{algorithmic}
\usepackage{algorithm}
\usepackage{array}
\usepackage{textcomp}
\usepackage{stfloats}
\usepackage{url}
\usepackage{verbatim}
\usepackage{graphicx}
\usepackage{cite}
\usepackage{multicol}
\usepackage{multirow}
\usepackage{makecell}
\usepackage{threeparttable}
\usepackage{xcolor}
\usepackage[caption=false]{subfig}
\begin{document}
\bstctlcite{IEEEexample:BSTcontrol} 

\title{Lightweight Generative Image Semantic Communication over Packet Erasure Channels}

\author{Yufei~Bo, Haoshuo~Zhang, Meixia~Tao,~\IEEEmembership{Fellow,~IEEE,} Jing~He, and Xinming~Huang
\thanks{Y. Bo, H. Zhang, and M. Tao are with the School of Information Science and Electronic Engineering, Shanghai Jiao Tong University, Shanghai 200240, China (email: \{boyufei01, zhuiguang, mxtao\}@sjtu.edu.cn).}
\thanks{J. He and X. Huang are with the National Key Laboratory for Positioning, Navigation and Timing Technology, College of Electronic Science and Technology, National University of Defense Technology, Changsha 410073, China (email: jing\_hj95@nudt.edu.cn; hxm\_kd@163.com).}}



\maketitle

\begin{abstract}
This paper addresses packet loss in semantic communication caused by network congestion or channel fluctuations. We propose LGSemCom, a lightweight generative packet-level joint source-channel coding (JSCC) framework for efficient and robust image transmission over packet erasure channels. Unlike conventional distortion-oriented recovery methods that yield blurry averages over erased regions, LGSemCom reformulates image recovery under packet erasures as a conditional generative task. By leveraging adversarial learning, the proposed decoder synthesizes plausible details without increasing complexity during inference. A key innovation of our framework is an erasure-aware weighting (EAW) strategy, which prioritizes generation in erased regions while preserving pixel fidelity in correctly received areas. To ensure computational efficiency, LGSemCom employs a fully convolutional codec based on efficient long-range attention blocks (ELABs) that capture global semantic dependencies with low complexity. Extensive experiments show that LGSemCom achieves superior perceptual reconstruction quality compared with existing benchmarks under severe packet loss, while achieving an order of magnitude faster inference. These attributes make LGSemCom highly suitable for latency-sensitive applications on resource-constrained edge devices. 
\end{abstract}


\section{Introduction}

\IEEEPARstart{S}{emantic} communication has emerged as a promising paradigm, particularly for challenging wireless or resource-constrained environments where reliably transmitting all source bits is either inefficient or infeasible.
By extracting and transmitting only task-relevant information, it offers a viable solution for scenarios like satellite communications with limited bandwidth and severe fading. 
As a typical end-to-end design of semantic communication, deep joint source-channel coding (JSCC) has been extensively studied to map source data to semantic features and transmit them directly over physical channels~\cite{deepJSCC, yang2023witt}.
However, these designs are incompatible with modern digital devices and protocols, hindering their practical deployment.
Recent efforts have shifted toward bit-level JSCC to enable semantic transmission over digital systems while preserving the benefits of end-to-end learning~\cite{zhang2025bitsemcom}.

While bit-level JSCC facilitates compatibility with digital physical-layer infrastructure, the resulting bitstreams are still transmitted in packets as required by standard communication pipelines and thus remain vulnerable to packet loss, which may be caused by network congestion or channel fluctuation. 
Conventional countermeasures, such as automatic repeat request (ARQ) and forward error correction (FEC), either introduce significant latency, making them unsuitable for delay-sensitive applications, or incur additional bandwidth overhead, which is prohibitive in resource-constrained environments.
To overcome this, recent studies have investigated packet-level JSCC to enhance resilience during packet transmission\cite{cheng2024grace, tian2025synchronous, wang2025resicomp}.
For instance, GRACE~\cite{cheng2024grace} jointly trains a pair of neural video encoder and decoder under simulated packet loss conditions.
ResiComp~\cite{wang2025resicomp} adopts masked visual token modeling to perform token prediction for packet loss concealment.
Despite their promise, these methods often rely on computationally intensive neural network (NN) architectures.
Moreover, they typically optimize for pixel-wise distortion metrics, which inevitably yield a blurry average over erased regions.
Consequently, achieving both high perceptual quality semantic reconstruction and low computational complexity under packet erasures remains an open challenge.

In this paper, we propose LGSemCom, a lightweight generative packet-level semantic communication system for efficient and robust image reconstruction over packet erasure channels.
Departing from conventional distortion-oriented recovery, LGSemCom reformulates image reconstruction under packet erasures as a conditional generative task. By integrating adversarial learning into training, the system synthesizes semantically consistent textures in erased regions without increasing computational complexity during inference.
A key innovation is the proposed erasure-aware weighting (EAW) strategy, which prioritizes generation in erased regions while preserving pixel fidelity in correctly received areas.
To ensure stable convergence under stochastic packet erasures, we further devise a multi-stage training strategy.
Architecturally, LGSemCom employs a fully convolutional codec based on efficient long-range attention blocks (ELABs) that capture global semantic dependencies with significantly reduced complexity.

\begin{figure*}[t]
    \centering
    \includegraphics[scale=0.52]{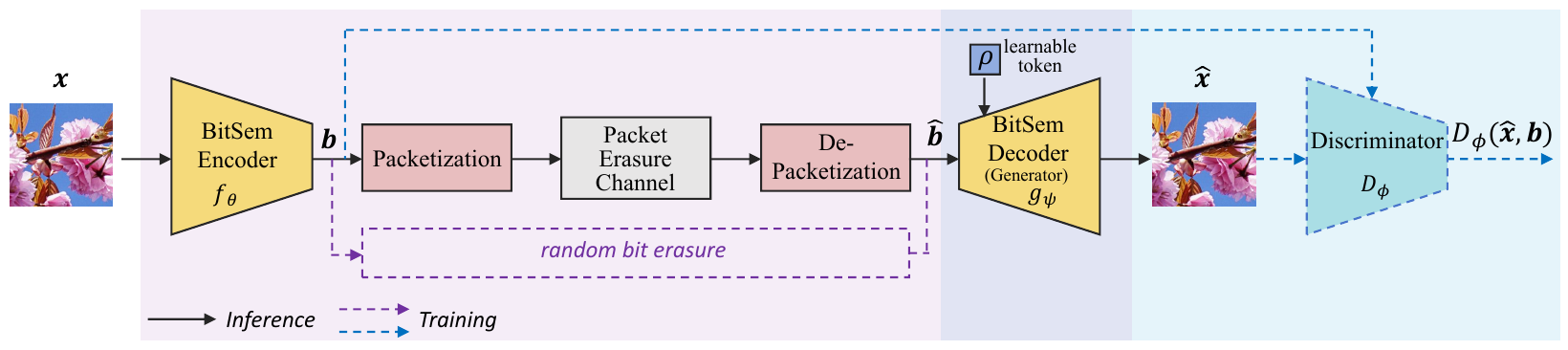}
    \vspace{-0.3cm}
    \caption{System model.}
    \label{system model}
    \vspace{-0.6cm}
\end{figure*}

Experimental results demonstrate that LGSemCom  outperforms both conventional FEC-based schemes and state-of-the-art (SOTA) learning-based benchmarks, particularly in terms of perceptual quality.
Ablation studies further show that EAW yields perceptual gains with negligible impact on fidelity. 
Notably, the proposed scheme facilitates reliable reconstruction without the need for retransmissions while achieving an order of magnitude faster inference.

\vspace{-0.2cm}
\section{Overall Architecture}

\subsection{System Model}
\vspace{-0.1cm}

We consider a generative semantic communication system designed for robust image transmission over packet erasure channels.
The overall framework is illustrated in Fig.~\ref{system model}.
The source image $\mathbf{x} \in \mathbb{R}^{H \times W \times C}$ is assumed to be a sample drawn from a natural image distribution $P_X$. 
At the transmitter, a bit-level semantic (BitSem) encoder $f_{\boldsymbol{\theta}}(\cdot)$ extracts semantic features and transforms them into a binarized feature map $\mathbf{b}\in\{0,1\}^{H_{\text{out}}\times W_{\text{out}}\times C_{\text{out}}}$ via a learnable bit mapper \cite{zhang2025bitsemcom}.
The resulting bit rate is defined as $r = \frac{H_{\text{out}} \times W_{\text{out}} \times C_{\text{out}}}{H \times W}$ bits per pixel (bpp).
To enhance resilience against bursty channel impairments, an interleaver-based packetization process is adopted to evenly distribute bits across the feature map into $N$ packets. 
These packets are transmitted over a packet erasure channel where random or bursty drops occur due to congestion or fading.
The receiver is assumed to accurately detect any dropped packets.
Upon de-packetization, the receiver identifies the lost packet indices and replaces the corresponding missing bits in the received sequence $\hat{\mathbf{b}}$ with a learnable token $\boldsymbol{\rho}$ by following the established masked token modeling \cite{he2022masked}.
The BitSem decoder $g_{\boldsymbol{\psi}}(\cdot)$ performs semantic-aware reconstruction directly from these inputs $\hat{\mathbf{x}}=g_{\boldsymbol{\psi}}(\hat{\mathbf{b}}, \boldsymbol{\rho})$, allowing for high quality inference from partially received data.

Unlike the traditional design goal of minimizing pixel-wise distortion metrics (\emph{e.g.}, mean square error (MSE) between $\mathbf{x}$ and $\hat{\mathbf{x}}$), our goal is to balance fidelity and perceptual consistency. We formulate the problem as
{\setlength{\abovedisplayskip}{5pt}
\setlength{\belowdisplayskip}{5pt}
\begin{align}
\mathbf{P1:}\quad &\min_{\boldsymbol{\theta}, \boldsymbol{\psi},\boldsymbol{\rho}} \quad \mathbb{E}_{\mathbf{x} \sim P_X, \mathbf{E} \sim P_E} \left[ d(\mathbf{x}, \hat{\mathbf{x}}) \right],\\
&\ \text{s.t.} \quad d_p(P_X, P_{\hat{X}}) \le \zeta,
\label{obj}\end{align}}where $\mathbf{E}$ denotes the random packet erasures generated by the channel distribution $P_E$, $d(\cdot, \cdot)$ quantifies fidelity distortion, $d_p(\cdot, \cdot)$ denotes the divergence between real and reconstructed image distributions, and $\zeta$ is a prescribed tolerance.

To enforce the perceptual constraint \eqref{obj}, we adopt a generative adversarial network (GAN)-based formulation by introducing a discriminator $D_{\boldsymbol{\phi}}(\cdot)$ in training.
The discriminator guides the generative decoder to synthesize semantically consistent textures in erased regions without increasing computational burden during deployment.

\vspace{-0.4cm}
\subsection{Network Architecture}

\begin{figure}[t]
  \centering
  \subfloat[Efficient long-range attention block (ELAB)]{
    \includegraphics[width=0.85\linewidth]{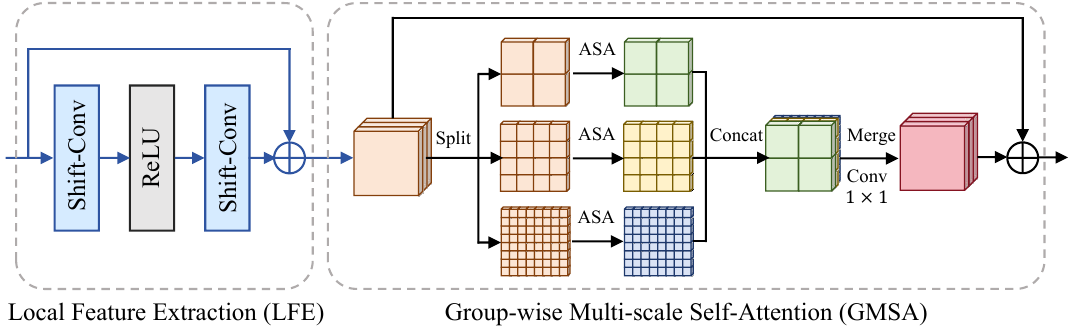}
  } \\ 
  \vspace{-0.3 cm}
  \subfloat[NN architecture of the lightweight BitSem codec]{
    \includegraphics[width=0.85\linewidth]{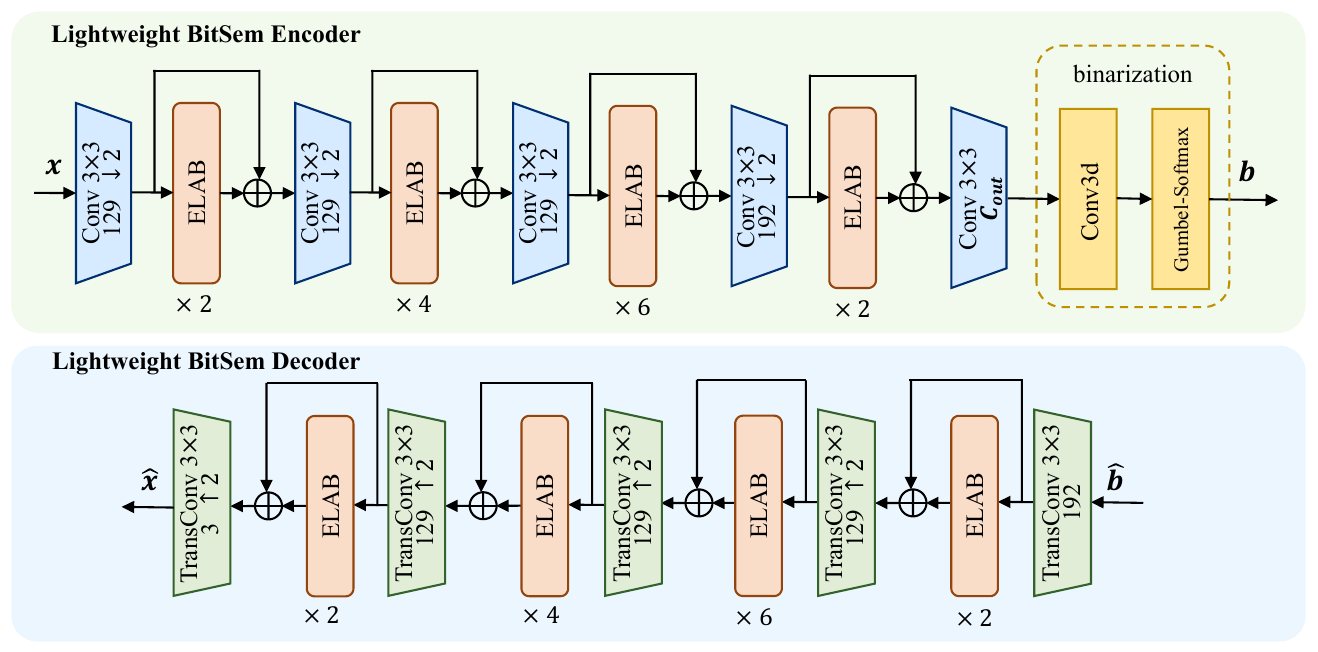}
  } \\
\vspace{-0.3 cm}
  \subfloat[NN architecture of the discriminator]{
    \includegraphics[width=0.85\linewidth]{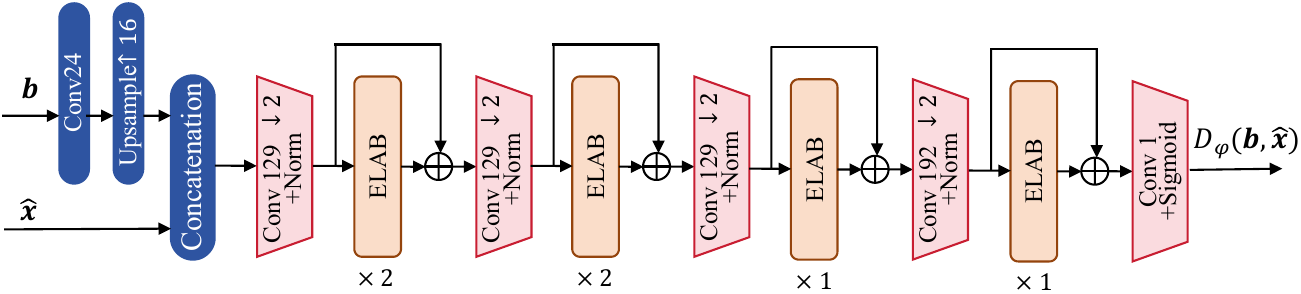}
  }
  \caption{NN architectures of the codec.}
  \label{codec}
  \vspace{-0.7 cm}
\end{figure}

To balance semantic extraction ability and computational efficiency, LGSemCom utilizes ELABs \cite{elab} as its backbone.
As shown in Fig.~\ref{codec}(a), ELAB achieves efficient feature extraction using local feature extraction (LFE) and group-wise multi-scale self-attention (GMSA) modules.
By using shift convolutions and accelerated self-attention (ASA), the codec captures long-range dependencies with reduced parameters.


The architecture of the BitSem codec is detailed in Fig.~\ref{codec}(b).
The encoder consists of four cascaded stages of ELABs with $3 \times3$ convolutional layers, which progressively downsample the input into a compact latent representation $\mathbf{m}\in \mathbb{R}^{H_{\text{out}}\times W_{\text{out}}\times C_{\text{out}}}$, where $H_{\text{out}} = H/16$ and $W_{\text{out}} = W/16$.
The learnable bit mapper \cite{zhang2025bitsemcom} is then used to transform the features into a discrete bit sequence $\mathbf{b} \in \{0, 1\}^{H_{\text{out}}\times W_{\text{out}}\times C_{\text{out}}}$, using a 3D convolutional layer to compute the probability of each bit being 0 or 1, and Gumbel-Softmax for differentiable sampling.
The BitSem decoder employs a symmetrical architecture, utilizing ELABs and transposed convolutions to reconstruct the source image $\hat{\mathbf{x}}$ from the received bits. 

The discriminator $D_{\boldsymbol{\phi}}$ is based on PatchGAN \cite{li2016precomputed}, which penalizes semantic similarity at the patch level, as shown in Fig.~\ref{codec}(c). 
Following the conditional GAN framework \cite{mentzer2020high}, $D_{\boldsymbol{\phi}}$ takes the decoder output $\hat{\mathbf{x}}$ as input, conditioned on the bit sequence $\mathbf{b}$.
Specifically, $\mathbf{b}$ is processed through a convolutional layer and upsampled to align with the dimensions of $\hat{\mathbf{x}}$. 
The concatenated features then pass through a series of downsampling stages consisting of convolutional layers, spectral normalization layers and ELABs. 
Finally, a $1 \times 1$ convolution with a Sigmoid activation is applied to produce the patch-level probability map $D_{\boldsymbol{\phi}}(\hat{\mathbf{x}},\mathbf{b}) \in (0, 1)^{H_{\text{out}}\times W_{\text{out}}}$.

\vspace{-0.1 cm}
\section{Loss Functions and Training Strategy}

\subsection{Training Objective}

To obtain a tractable learning objective from the constrained optimization problem $\mathbf{P1}$, we adopt a Lagrangian-like relaxation by converting the perceptual constraint into an adversarial loss within the objective.  
Specifically, the perceptual constraint \eqref{obj} is enforced as a GAN-based min-max game, and the distortion $d(\mathbf{x}, \hat{\mathbf{x}})$ is quantified by MSE, yielding the objective:
{\setlength{\abovedisplayskip}{4pt}
\setlength{\belowdisplayskip}{4pt}
\begin{equation}
\label{minmax}
\min_{\boldsymbol{\theta}, \boldsymbol{\psi}, \boldsymbol{\rho}} \max_{\boldsymbol{\phi}} \quad \mathbb{E}_{\mathbf{x},\mathbf{E}} [\|\mathbf{x} - \hat{\mathbf{x}}\|_2^2] + \lambda \mathcal{L}_{\text{GAN}}(\mathbf{x}, \hat{\mathbf{x}}, D_{\boldsymbol{\phi}}).
\end{equation}}Here, $\|\cdot\|_2$ denotes the $l_2$ norm, $\lambda$ is the hyperparameter that trades off perception with distortion, and $\mathcal{L}_{\text{GAN}}(\mathbf{x}, \hat{\mathbf{x}}, D_{\boldsymbol{\phi}})$ unifies the adversarial objectives for both the generator and the discriminator, as detailed later.

\vspace{-0.3cm}
\subsection{Multi-Stage Training Strategy}

Directly training the network via \eqref{minmax} often suffers from instability, as an uninitialized generator tends to produce severe artifacts, which causes the discriminator to dominate and consequently leads to vanishing adversarial gradients.
To ensure stable convergence, LGSemCom adopts a multi-stage strategy that progressively builds erasure resilience before introducing adversarial learning.

\subsubsection{Backbone pre-training}

The first stage pre-trains the BitSem codec in a lossless environment using the MSE loss.
This stage provides a strong initialization for semantic feature extraction and reconstruction.

\subsubsection{Joint optimization for erasure resilience}

The second stage introduces the packet erasure channel for end-to-end optimization.
Since our packetization distributes bits uniformly across $N$ packets, packet erasures are statistically equivalent to independent and identically distributed (i.i.d.) bit-level erasures in $\mathbf{b}$.
Therefore, we simulate varying packet loss rates by randomly masking $k\%$ of the bits per iteration, with $k \sim \mathcal{U}[0, k_{\max}]$ and $k_{\max} = 87.5$.
The loss function is
{\setlength{\abovedisplayskip}{4pt}
\setlength{\belowdisplayskip}{4pt}
\begin{equation}
\small
\mathcal{L}_{2}(\boldsymbol{\theta}, \boldsymbol{\psi}, \boldsymbol{\rho}) = \mathbb{E}_{\mathbf{x}, \mathbf{E}} \left[ \| \mathbf{x} - g_{\boldsymbol{\psi}}(\mathcal{M}(f_{\boldsymbol{\theta}}(\mathbf{x}), \mathbf{E}), \boldsymbol{\rho}) \|_2^2 \right],
\end{equation}}where $\mathcal{M}(\cdot, \mathbf{E})$ denotes the erasure operator given $\mathbf{E}$.  This stage equips the codec with robustness to stochastic erasures while preserving pixel fidelity.

\subsubsection{Adversarial training}

The final stage introduces adversarial training to enhance perceptual quality.
To solve the min-max optimization in \eqref{minmax}, we iteratively update the generative decoder $g_{\boldsymbol{\psi}}$ and discriminator $D_{\boldsymbol{\phi}}$ while keeping the encoder frozen to maintain latent-space stability.
Their loss functions are defined as follows.

In each training iteration, the patch-based discriminator $D_{\boldsymbol{\phi}}$ is updated first to distinguish real samples $\mathbf{x}$ from reconstructed samples $\hat{\mathbf{x}}$.
Specifically, we first define an element-wise binary cross-entropy loss matrix $\mathbf{L}_{D_{\text{std}}} \in \mathbb{R}^{H_{\text{out}} \times W_{\text{out}}}$:
{\setlength{\abovedisplayskip}{5pt}
\setlength{\belowdisplayskip}{5pt}
\begin{equation}
\small
\mathbf{L}_{D_{\text{std}}} = -\left( \log D_{\boldsymbol{\phi}}(\mathbf{x}, \mathbf{b}) + \log(1 - D_{\boldsymbol{\phi}}(\hat{\mathbf{x}}, \mathbf{b})) \right),
\end{equation}}where $\log$ is applied element-wise  across all spatial coordinates $(i,j)$, with $i \in \{1, \dots, H_{\text{out}}\}$ and $j \in \{1, \dots, W_{\text{out}}\}$. Then, the standard discriminator loss $\mathcal{L}_{D_{\text{std}}}$ is optimized by averaging $\mathbf{L}_{D_{\text{std}}}$ over all coordinates
{\setlength{\abovedisplayskip}{5pt}
\setlength{\belowdisplayskip}{5pt}
\begin{equation}
\small
\mathcal{L}_{D_{\text{std}}} = \mathbb{E}_{\mathbf{x},\mathbf{E}} \left[ \frac{1}{H_{\text{out}}W_{\text{out}}} \sum\nolimits_{i,j} \left( \mathbf{L}_{D_{\text{std}}} \right)_{i,j} \right].
\end{equation}}Following the discriminator update, the generative decoder is optimized to synthesize realistic details to ``fool'' the discriminator.
A standard adversarial objective would directly apply a uniform perceptual penalty across the entire image as:
{\setlength{\abovedisplayskip}{5pt}
\setlength{\belowdisplayskip}{5pt}
\begin{equation}
\small
\mathcal{L}_{G_{\text{std}}} = \mathbb{E}_{\mathbf{x},\mathbf{E}} \left[ \Vert{}\mathbf{x} - \hat{\mathbf{x}}\Vert{}_2^2 - \frac{\lambda}{H_{\text{out}}W_{\text{out}}} \sum_{i,j} \log(D_{\boldsymbol{\phi}}(\hat{\mathbf{x}}, \mathbf{b}))_{i,j} \right].
\end{equation}}

However, uniform weighting ignores that the received and erased regions have different recovery priorities.
In received regions, the objective should remain distortion-centric to ensure fidelity.
Conversely, in erased regions, perceptually driven generation is required to avoid regression-to-mean blurring.



\begin{figure*}[t]
  \centering
  \subfloat[PSNR vs. PLR \label{msssim}]{
    \includegraphics[width=0.30\linewidth]{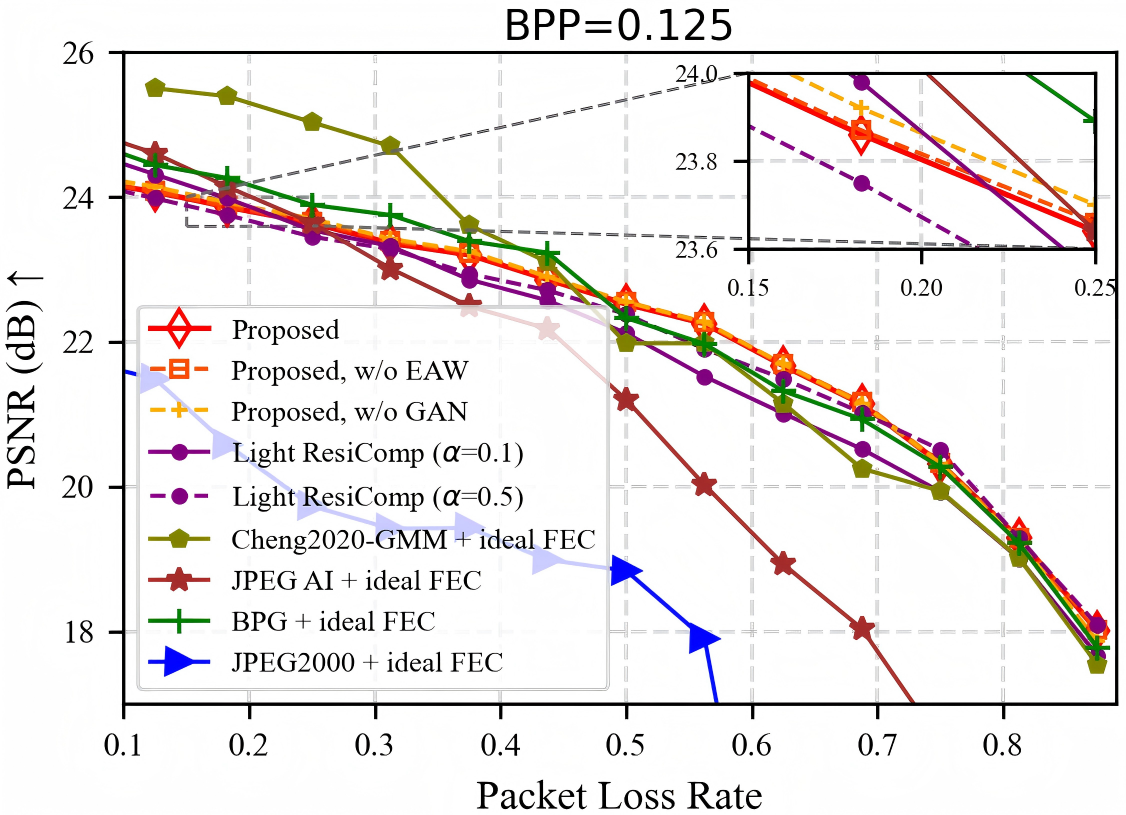}
  }
  \hfill 
  \subfloat[LPIPS vs. PLR \label{feature}]{
    \includegraphics[width=0.30\linewidth]{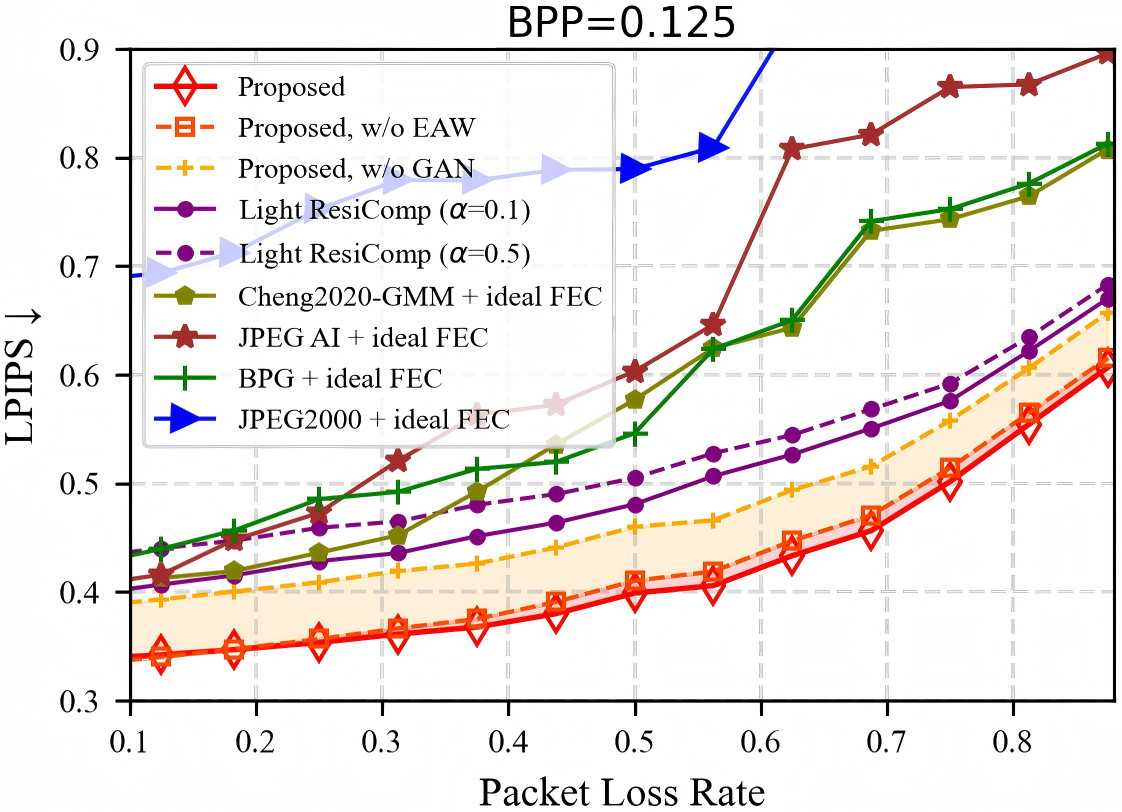}
  }
  \hfill
  \subfloat[FID vs. PLR \label{rate}]{
    \includegraphics[width=0.30\linewidth]{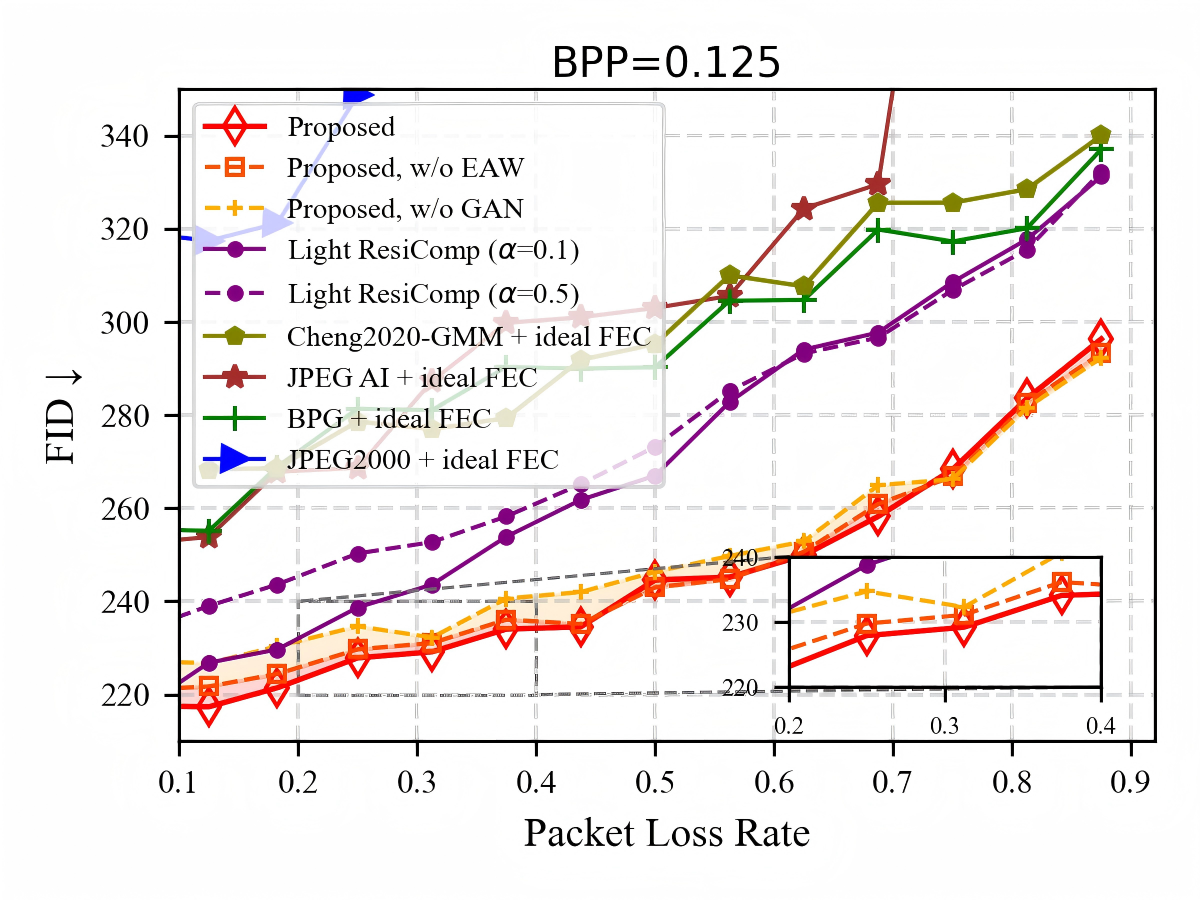}
  }

  \vspace{-0.1cm} 
  \caption{Performance comparison under varying PLRs across different metrics at $bpp=0.125$.}
  \label{results}
  \vspace{-0.8cm} 
\end{figure*}

\begin{figure}[t]
  \centering
  \subfloat[PSNR vs. bpp \label{msssim}]{
    \includegraphics[width=0.47\linewidth]{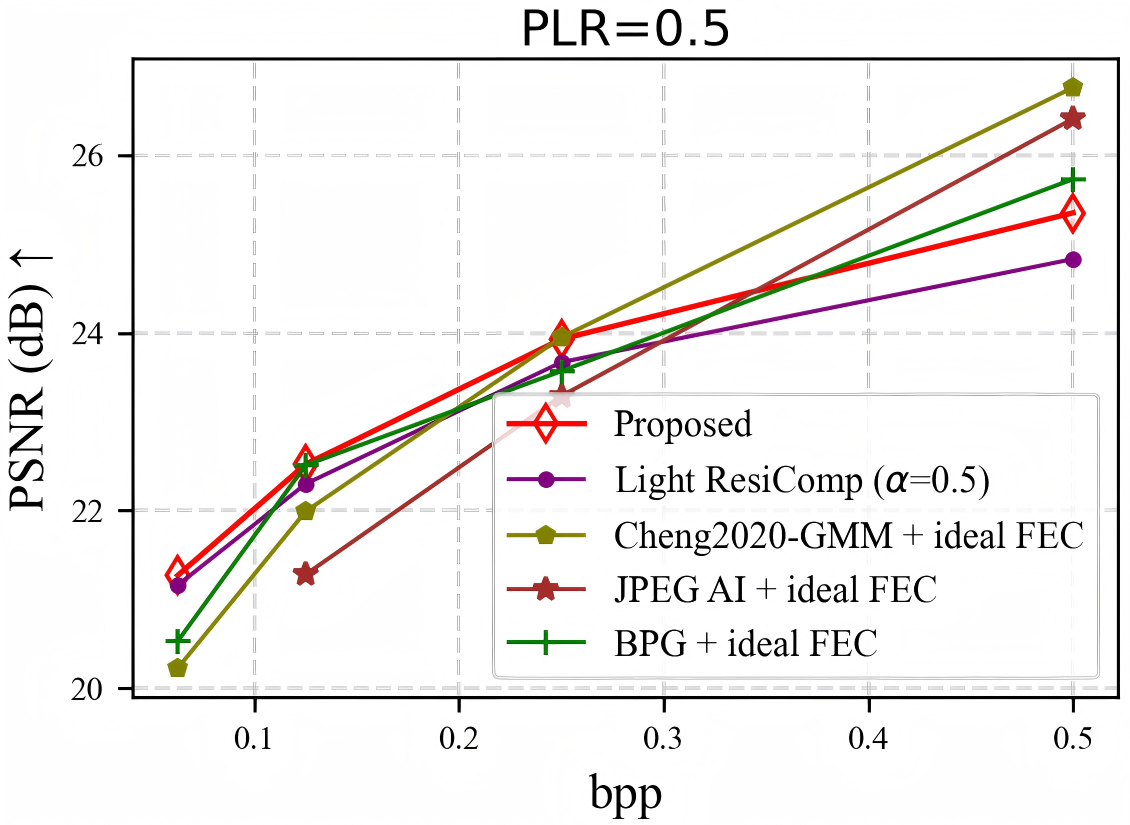}
  }
  \hfill 
  \subfloat[LPIPS vs. bpp \label{feature}]{
    \includegraphics[width=0.47\linewidth]{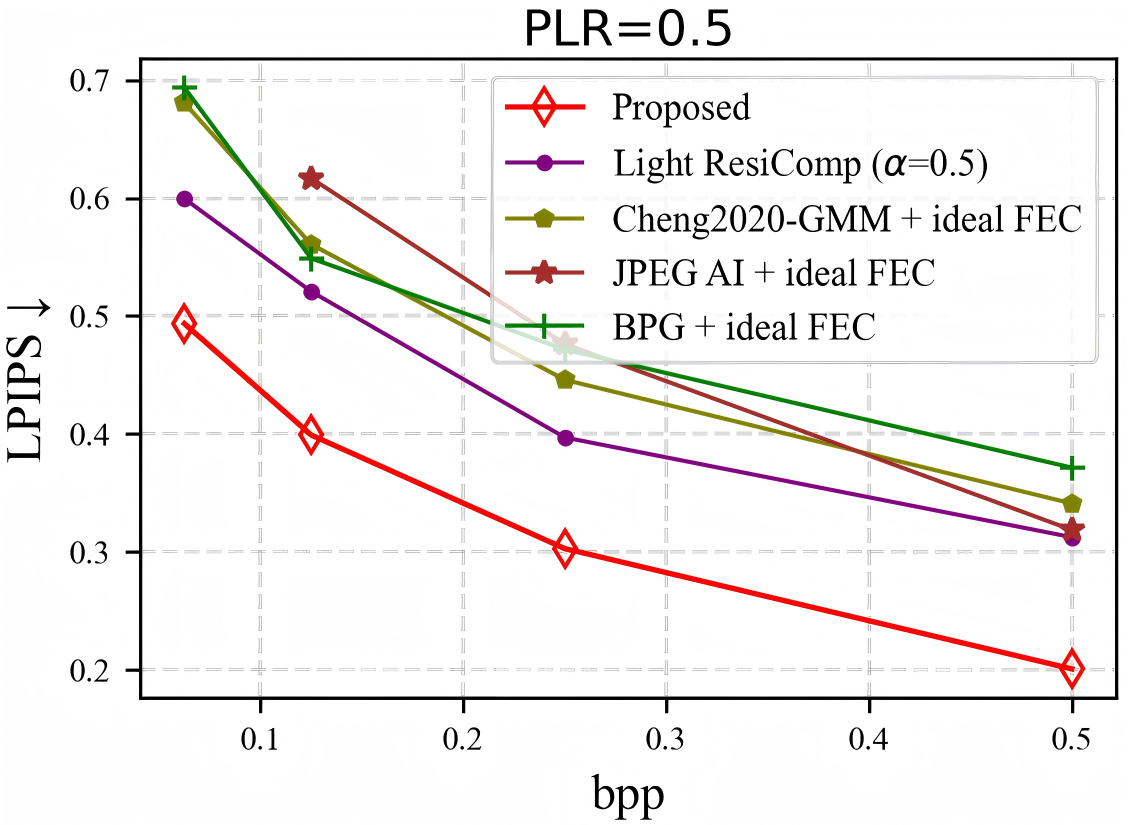}
  }
  \vspace{-0.1cm} 
  \caption{Performance comparison under varying bpps at $\text{PLR}=0.5$.}
  \label{r-d}
  \vspace{-0.3cm} 
\end{figure}



\begin{figure}[t]
    \centering
    \includegraphics[scale=0.35]{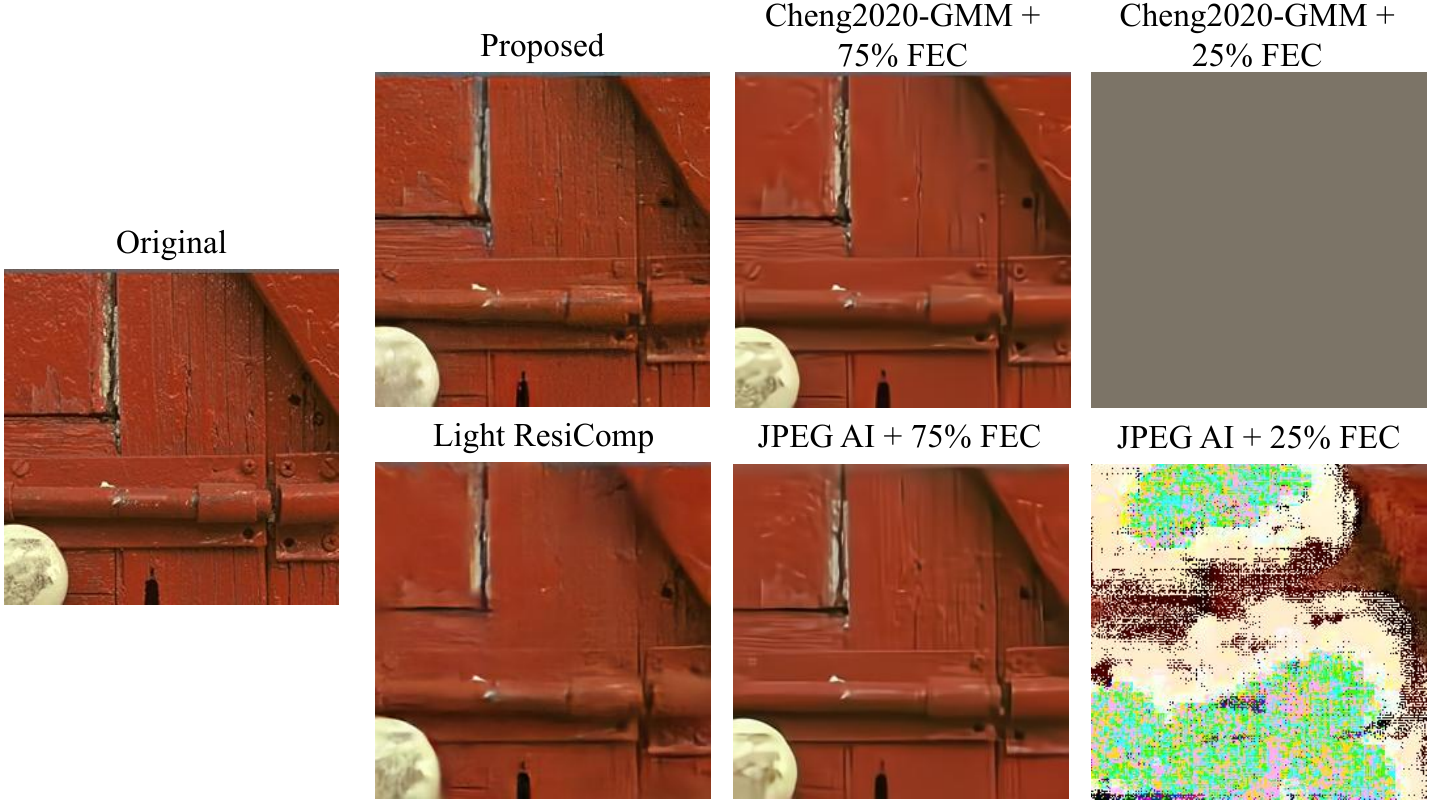}
    \vspace{-0.3cm}
    \caption{Visualization of reconstructed images with $8$ out of $16$ packets lost.}
    \label{visual_comparison}
    \vspace{-0.5cm}
\end{figure}

To achieve a spatially-varying balance between pixel fidelity and perceptual quality, we propose the EAW strategy.
Let $\mathbf{E} \in \{0, 1\}^{H_{\text{out}} \times W_{\text{out}}}$ denote the binary erasure mask, where $E_{i,j} = 1$ indicates bit loss at coordinate $(i,j)$ of the feature map.
The weight matrix $\mathbf{M}$ is constructed as
{\setlength{\abovedisplayskip}{5pt}
\setlength{\belowdisplayskip}{5pt}
\begin{equation}
\small
\mathbf{M} = \mathbf{1} + \omega \cdot \mathbf{E},
\end{equation}}where $\omega > 0$ determines the intensity of the perceptual synthesis in the lost region. 
The erasure-aware generator objective replaces the uniform penalty with spatially-varying weighting:
{\setlength{\abovedisplayskip}{5pt}
\setlength{\belowdisplayskip}{5pt}
\begin{equation}
\small
    \mathcal{L}_{G_{\text{EAW}}}\!=\!\mathbb{E}_{\mathbf{x},\mathbf{E}} \!\left[ \Vert{}\mathbf{x} - \hat{\mathbf{x}}\Vert{}_2^2 \!- \!\frac{\lambda}{H_{\text{out}}W_{\text{out}}}\! \sum_{i,j}\! \left( \mathbf{M} \odot \log D_{\boldsymbol{\phi}}(\hat{\mathbf{x}}, \mathbf{b}) \right)_{i,j} \right]\!,
\end{equation}}where $\odot$ represents the element-wise product.

To ensure semantic consistency, we further integrate a multi-layer feature matching (FM) loss\cite{wang2018high}, which suppresses generative artifacts by matching intermediate representations of $D_{\boldsymbol{\phi}}$ from real and reconstructed images:
{\setlength{\abovedisplayskip}{4pt}
\setlength{\belowdisplayskip}{4pt}
\begin{equation}
\small
\mathcal{L}_{\text{FM}} = \sum_{i=1}^{L} \frac{1}{N_i} \left\| D_{\boldsymbol{\phi}}^{(i)}(\mathbf{x}, \mathbf{b}) - D_{\boldsymbol{\phi}}^{(i)}(\hat{\mathbf{x}}, \mathbf{b}) \right\|_1,
\end{equation}}where $\|\cdot\|_1$ is the $l_1$ norm, $D_{\boldsymbol{\phi}}^{(i)}$ denotes the feature map of the $i$-th layer (out of $L$) and $N_i$ is its total element count.

Combining these components, the min-max objective in \eqref{minmax} is decomposed into two loss functions for alternately updating the generative decoder $g_{\boldsymbol{\psi}}$ and the discriminator $D_{\boldsymbol{\phi}}$ as:
{\setlength{\abovedisplayskip}{4pt}
\setlength{\belowdisplayskip}{4pt}
\begin{align}
\small
\mathcal{L}_{G} &= \mathcal{L}_{G_{\text{EAW}}} + \sigma \mathcal{L}_{FM},\\
\mathcal{L}_D &= \mathbb{E}_{\mathbf{x},\mathbf{E}} \left[ \frac{1}{H_{\text{out}}W_{\text{out}}} \sum\nolimits_{i,j} \left( \mathbf{M} \odot \mathbf{L}_{D_{\text{std}}} \right)_{i,j} \right],
\end{align}}where $\sigma$ is a hyperparameter.


\vspace{-0.3cm}
\section{Experimental Results}

\subsection{Experimental Settings}
\vspace{-0.1cm}

LGSemCom is trained and evaluated on DIV2K~\cite{agustsson2017ntire}.
The dataset is split into training, validation, and testing sets with a ratio of 8:1:1. 
During training, images are randomly cropped to patches of size $256 \times 256$, while during testing, center cropping is applied.
We use the Adam optimizer with a batch size of 4.
The number of packets, $N$, is set to 16, and the channel output dimension $C_{\text{out}}$ is adjusted to control the bit rate.
The coefficients $\lambda$, $\omega$, and $\sigma$ are set to 1, 2 and 10, respectively.

We compare LGSemCom with two types of benchmarks: 

\textbf{1) Light ResiComp: }
ResiComp\cite{wang2025resicomp} enhances the resilience of neural image codecs by incorporating masked visual token modeling from large vision models. 
To ensure a fair comparison using the same backbone architecture, we implement a ``Light ResiComp'' by replacing its heavy Transformer blocks with ELABs. This reduces its total model parameters from 128M to 8.4M while retaining its core coding mechanism.
Additionally, ResiComp introduces a parameter, $\alpha$, to control the tradeoff between reconstruction performance under perfect transmission and resilience to packet loss.
A larger value of $\alpha$ places more emphasis on resilience.
In our simulations, we evaluate $\alpha = 0.1$ and $\alpha = 0.5$.

\textbf{2) Image compression with ideal packet-level FEC:} 
These methods adopt a modular design that separates source compression and error protection.
The considered image compression algorithms include two learning-based methods, JPEG AI\cite{ascenso2023jpeg} and the Gaussian mixture model-based scheme \cite{cheng2020learned} (referred to as Cheng2020-GMM hereinafter), and the conventional BPG and JPEG2000.
For error protection, we employ ideal FEC with parity packet ratios of $\{12.5\%, 25\%, 37.5\%, 50\%, 62.5\%, 75\%, 87.5\%\}$ and plot the envelope of the results.
Here, the parity packet ratio is defined as the proportion of parity packets $N_r$ to the total number of packets $N$.
An image is considered successfully decoded at the receiver if no more than $N_r$ packets out of $N$ are lost. 
In cases where decoding fails, a random image is generated to represent the unsuccessful reconstruction.


To evaluate reconstruction quality, we adopt three metrics: PSNR for pixel-wise fidelity, LPIPS\cite{zhang2018unreasonable} for perceptual quality, and Fréchet Inception Distance (FID)\cite{heusel2017gans} for distributional distance.


\begin{table}[t]
\renewcommand{\arraystretch}{1.1}
\setlength{\tabcolsep}{3.5 pt}
\centering
\begin{threeparttable}[b]
\caption{Performance under bursty packet loss scenarios}
\vspace{-0.1cm}
\begin{tabular}{cccccc}
\hline
Scenario             & Method      & PSNR$\uparrow$           & LPIPS$\downarrow$           & FID$\downarrow$                       \\ \hline
\multirow{8}{*}{\makecell{M \\ (PLR$=13.8\%$\\$\gamma=1.69$)}} & BPG+25\%FEC & 25.71          & 0.3615          & 177.37                  \\
                     & BPG+50\%FEC & 28.51 & 0.2376          & 157.86  
                     \\ & JPEG AI+25\%FEC & 25.51 & 0.3357          & 154.53      \\
                     & JPEG AI+50\%FEC & 28.40 & 0.2219          & 150.65      \\
                     & Cheng2020-GMM+25\%FEC & 25.81 & 0.3677          & 195.33      \\
                     & Cheng2020-GMM+50\%FEC & \textbf{29.10} & 0.2250         & 184.66      \\
                     & Light ResiComp & 27.20           & 0.2075          & 146.56          \\
                     & Proposed    & 27.22          & \textbf{0.1702} & \textbf{129.37}  \\ \hline
\multirow{8}{*}{\makecell{S \\ (PLR$=32.3\%$\\$\gamma=2.71$)}} & BPG+25\%FEC & 18.12          & 0.6750           & 301.36                 \\
                     & BPG+50\%FEC & 25.74 & 0.3581          & 184.74                 \\
                     & JPEG AI+25\%FEC & 18.77 & 0.6257          & 257.58      \\
                     & JPEG AI+50\%FEC & 25.57 & 0.3467          & 177.85      \\
                     & Cheng2020-GMM+25\%FEC & 18.93 & 0.6476          & 294.83      \\
                     & Cheng2020-GMM+50\%FEC & 25.96 & 0.3697          & 210.98      \\
                     & Light ResiComp & 25.48          & 0.2692          & 188.52           \\
                     & Proposed    & \textbf{26.19} & \textbf{0.1931} & \textbf{146.58}  \\ \hline
\end{tabular}
   \label{real}
  \end{threeparttable}
  \vspace{-0.5cm}
\end{table}

\vspace{-0.3cm}
\subsection{Performance under i.i.d. Packet Loss}

We first consider i.i.d. packet loss scenarios. 
Fig.~\ref{results}(a) illustrates the PSNR performance under varying packet loss rates (PLRs) at $bpp=0.125$. 
Although some separation schemes, particularly Cheng2020-GMM and BPG with ideal FEC, yield higher PSNR under mild packet loss ($\text{PLR} < 0.45$), their performance degrades rapidly as the PLR increases. In contrast, the proposed method demonstrates superior robustness in high PLR scenarios, eventually matching and then surpassing these benchmarks.
Moreover, the proposed method generally outperforms Light ResiComp. 
Specifically, it achieves a higher PSNR than Light ResiComp with $\alpha = 0.5$. 
While the less resilient variant ($\alpha = 0.1$) performs better in low PLR regions, its performance degrades in high PLR regions.
This gain over Light ResiComp is attributed to the encoding strategy. 
The entropy-based encoding in Light ResiComp produces discrete tokens of varying importance, where the loss of critical tokens greatly degrades performance.
In contrast, our method combines end-to-end bit mapping with an interleaver.
This design distributes information evenly across the bits, thus improving resilience to packet loss.

Perceptual performance, in terms of LPIPS and FID, is shown in Fig.~\ref{results}(b) and (c), respectively.
LGSemCom consistently outperforms all benchmarks across all PLRs.
Furthermore, the introduction of GAN significantly enhances perceptual quality with only a marginal degradation in PSNR, as shown in Fig.~\ref{results}(a).
Further integrating EAW provides additional perceptual gains, reducing LPIPS and FID by up to 0.02 and 3, respectively, with negligible impact on PSNR. 
These results indicate that EAW effectively concentrates generation on the lost regions without compromising pixel fidelity in correctly received regions.

Fig.~\ref{r-d} further illustrates the rate-distortion performance at $\text{PLR} = 0.5$. The proposed framework demonstrates superiority in both PSNR and LPIPS, particularly under low-to-medium bitrates.
A visual comparison of the reconstructed images with 8 out of 16 packets lost at $bpp=0.5$ is illustrated in Fig. \ref{visual_comparison}.
As can be seen, while conventional separation schemes exhibit either quality degradation or total collapse, our method preserves sharp visual details and superior perceptual quality. 

\begin{table}[t]
\renewcommand{\arraystretch}{1.1}
\setlength{\tabcolsep}{1.7pt}
\centering
\begin{threeparttable}[b]
\caption{Computational complexity and inference time}
\vspace{-0.1cm}
\begin{tabular}{ccccc}
\hline
         & \#Params(M) & FLOPs(G) & Enc. Time(ms)$^\star$       & Dec. Time(ms)$^\star$ \\ \hline
Proposed & 6.15         & 18.06     & 11.2                 & 11.3           \\
Light ResiComp & 8.39         & 28.71     & 246.9                    & 249.7              \\
JPEG AI  & -            & -         & \textgreater{}$10^3$ & 266.6          \\ Cheng2020-GMM & 11.83         & 27.313    &  391.22                    & 886.91              \\
\hline
\end{tabular}
   \begin{tablenotes}
     \item[$^\star$] All experiments are conducted on an Intel Xeon Silver 4214R CPU, and a 48 GB Nvidia GeForce RTX 4090 Ti graphics card. Note that the parameter count and FLOPs for JPEG AI are omitted as its encapsulated software package prevents direct complexity measurement.
   \end{tablenotes}
     \label{complexity}
  \end{threeparttable}
\vspace{-0.8cm}
\end{table}

\vspace{-0.4cm}
\subsection{Performance under Realistic Packet Loss}

To evaluate performance under realistic packet loss conditions, we adopt a first-order, 3-state Markov chain-based model from \cite{wang2025resicomp}, which reproduces the packet loss traces of real mobile networks.
The model comprises a non-loss state ($S_G$), a lossy state ($S_B$), and an intermediary state ($S_I$).
Specifically, we select two representative cases corresponding to moderate (M) and severe (S) loss conditions, characterized by average PLRs of 13.8$\%$ and 32.3$\%$, and average burst lengths $\gamma$ of 1.69 and 2.71, respectively. 

Table~\ref{real} summarizes the results with $bpp=0.5$.
The proposed method shows significant advantages in perceptual metrics. 
In the M scenario, LGSemCom achieves the lowest LPIPS and FID, outperforming other baselines by a clear margin. Although Cheng2020-GMM with 50$\%$ FEC yields a higher PSNR, its perceptual quality degrades.
In the S scenario, LGSemCom achieves the best performance across all metrics, showing its robustness under realistic packet loss conditions.

\vspace{-0.4cm}
\subsection{Computational Complexity}

Table~\ref{complexity} compares the computational complexity and inference time of learning-based methods. 
The proposed method outperforms the others in both aspects, with the fewest parameters and FLOPs, and the fastest encoding time and decoding time.
In comparison, Light ResiComp requires 36.42$\%$ more parameters and 58.97$\%$ more FLOPs than the proposed method due to its entropy models. 
Its inference time is 20 times longer than the proposed method, resulting from its complex layered context modeling and entropy coding.
The proposed method also achieves a significantly faster processing speed compared to JPEG AI and Cheng2020-GMM.

\section{Conclusion}
This paper presents LGSemCom, a lightweight generative packet-level semantic communication system that enables semantic-aware image reconstruction from partially received packets.
By introducing an erasure-aware adversarial loss that generates plausible details in lost regions, LGSemCom reconstructs images with superior perceptual quality.
LGSemCom also achieves an order of magnitude faster inference, making it a promising solution for latency-sensitive applications.

While our present setting assumes perfect receiver packet loss detection, future efforts could address scenarios with imperfect detection to further enhance practical robustness.
Another promising direction is the extension of LGSemCom to additional modalities like video streaming and broader vision tasks.

\bibliographystyle{IEEEtran}
\bibliography{gpsemcom.bib}{}

\begin{thebibliography}{10}
\providecommand{\url}[1]{#1}
\csname url@samestyle\endcsname
\providecommand{\newblock}{\relax}
\providecommand{\bibinfo}[2]{#2}
\providecommand{\BIBentrySTDinterwordspacing}{\spaceskip=0pt\relax}
\providecommand{\BIBentryALTinterwordstretchfactor}{4}
\providecommand{\BIBentryALTinterwordspacing}{\spaceskip=\fontdimen2\font plus
\BIBentryALTinterwordstretchfactor\fontdimen3\font minus
  \fontdimen4\font\relax}
\providecommand{\BIBforeignlanguage}[2]{{%
\expandafter\ifx\csname l@#1\endcsname\relax
\typeout{** WARNING: IEEEtran.bst: No hyphenation pattern has been}%
\typeout{** loaded for the language `#1'. Using the pattern for}%
\typeout{** the default language instead.}%
\else
\language=\csname l@#1\endcsname
\fi
#2}}
\providecommand{\BIBdecl}{\relax}
\BIBdecl

\bibitem{deepJSCC}
E.~Bourtsoulatze, D.~Burth~Kurka, and D.~Gündüz, ``Deep joint source-channel
  coding for wireless image transmission,'' \emph{IEEE Trans. Cogn. Commun.
  Netw.}, vol.~5, no.~3, pp. 567--579, 2019.

\bibitem{yang2023witt}
K.~Yang \emph{et~al.}, ``{WITT}: A wireless image transmission transformer for
  semantic communications,'' in \emph{Proc. IEEE Int. Conf. Acoust., Speech
  Signal Process. (ICASSP)}, Rhodes Island, Greece, 2023, pp. 1--5.

\bibitem{zhang2025bitsemcom}
H.~Zhang, Y.~Bo, J.~Mo, and M.~Tao, ``{BitSemCom}: A bit-level semantic
  communication framework with learnable probabilistic mapping,'' \emph{IEEE
  Commun. Lett.}, vol.~30, pp. 2357--2361, 2026.

\bibitem{cheng2024grace}
Y.~Cheng \emph{et~al.}, ``{GRACE}:{Loss-Resilient Real-Time} video through
  neural codecs,'' in \emph{Proc. USENIX Symp. Netw. Syst. Design Implement.
  (NSDI)}, Santa Clara, CA, USA, 2024, pp. 509--531.

\bibitem{tian2025synchronous}
Y.~Tian, J.~Ying, Z.~Qin, Y.~Jin, and X.~Tao, ``Synchronous multi-modal
  semantic communication system with packet-level coding,'' \emph{IEEE Trans.
  Wireless Commun.}, vol.~24, no.~5, pp. 3684--3697, 2025.

\bibitem{wang2025resicomp}
S.~Wang \emph{et~al.}, ``{ResiComp}: Loss-resilient image compression via
  dual-functional masked visual token modeling,'' \emph{IEEE Trans. Circuits
  Syst. Video Technol.}, vol.~35, no.~7, pp. 7181--7195, 2025.

\bibitem{he2022masked}
K.~He, X.~Chen, S.~Xie, Y.~Li, P.~Doll{\'a}r, and R.~Girshick, ``Masked
  autoencoders are scalable vision learners,'' in \emph{Proc. IEEE/CVF Conf.
  Comput. Vis. Pattern Recogn. (CVPR)}, 2022, pp. 16\,000--16\,009.

\bibitem{elab}
X.~Zhang, H.~Zeng, S.~Guo, and L.~Zhang, ``Efficient long-range attention
  network for image super-resolution,'' in \emph{Proc. Eur. Conf. Comput. Vis.
  (ECCV)}, Tel Aviv, Israel, 2022, pp. 649--667.

\bibitem{li2016precomputed}
C.~Li and M.~Wand, ``Precomputed real-time texture synthesis with {{M}arkovian}
  generative adversarial networks,'' in \emph{Proc. Eur. Conf. Comput. Vis.
  (ECCV)}, Amsterdam, The Netherlands, 2016, pp. 702--716.

\bibitem{mentzer2020high}
F.~Mentzer \emph{et~al.}, ``High-fidelity generative image compression,''
  \emph{Adv. Neural Inf. Process. Syst. (NeurIPS)}, vol.~33, pp.
  11\,913--11\,924, 2020.

\bibitem{wang2018high}
T.-C. Wang \emph{et~al.}, ``High-resolution image synthesis and semantic
  manipulation with conditional {GANs},'' in \emph{Proc. IEEE Conf. Comput.
  Vis. Pattern Recognit. (CVPR)}, Salt Lake City, UT, USA, 2018, pp.
  8798--8807.

\bibitem{agustsson2017ntire}
E.~Agustsson and R.~Timofte, ``{{NTIRE}} 2017 challenge on single image
  super-resolution: Dataset and study,'' in \emph{Proc. IEEE Conf. Comput. Vis.
  Pattern Recognit. Workshops}, Honolulu, HI, USA, 2017, pp. 126--135.

\bibitem{ascenso2023jpeg}
J.~Ascenso, E.~Alshina, and T.~Ebrahimi, ``The {JPEG AI} standard: Providing
  efficient human and machine visual data consumption,'' \emph{IEEE Multimed.},
  vol.~30, no.~1, pp. 100--111, 2023.

\bibitem{cheng2020learned}
Z.~Cheng, H.~Sun, M.~Takeuchi, and J.~Katto, ``Learned image compression with
  discretized {Gaussian} mixture likelihoods and attention modules,'' in
  \emph{Proc. IEEE/CVF Conf. Comput. Vis. Pattern Recogn. (CVPR)}, 2020, pp.
  7939--7948.

\bibitem{zhang2018unreasonable}
R.~Zhang \emph{et~al.}, ``The unreasonable effectiveness of deep features as a
  perceptual metric,'' in \emph{Proc. IEEE Conf. Comput. Vis. Pattern Recognit.
  (CVPR)}, Salt Lake City, UT, USA, 2018, pp. 586--595.

\bibitem{heusel2017gans}
M.~Heusel \emph{et~al.}, ``{GAN}s trained by a two time-scale update rule
  converge to a local nash equilibrium,'' \emph{Adv. Neural Inf. Process. Syst.
  (NIPS)}, vol.~30, 2017.

\end{thebibliography}

\end{document}